\documentclass[preprint]{aastex62}

\newcommand{\logg} {\mbox{{\rm logg}}}
\newcommand{\feh} {\mbox{\rm [Fe/H]}}
\newcommand{\afe} {\mbox{\rm [$\alpha$/Fe]}}

\newcommand{\VR} {\mbox{\rm $V_{\rm{R}}$}}
\newcommand{\Vphi} {\mbox{\rm $V_{\rm{\phi}}$}}
\newcommand{\kmprs} {\mbox{\rm \,$km/s$\,}}

\newcommand{\Rapo} {\mbox{\rm $R_{\rm apo}$}}
\newcommand{\Rperi} {\mbox{\rm $R_{\rm peri}$}}
\newcommand{\Zmax} {\mbox{\rm $Z_{\rm max}$}}
\newcommand{\Rm} {\mbox{\rm $R_{\rm m}$}}
\newcommand{\Lz} {\mbox{\rm $L_{\rm z}$}}
\newcommand{\Etot} {\mbox{\rm $E_{\rm tot}$}}
\newcommand{\Etotn} {\mbox{\rm $E_{\rm tot}/{\rm 10^5}$}}

\begin{document}

\title{Low-$\alpha$ Metal-Rich Stars with Sausage Kinematics in the LAMOST Survey: Are they from the Gaia-Sausage-Enceladus Galaxy?}

\correspondingauthor{Gang Zhao}
\email{gzhao@nao.cas.cn}

\author[0000-0002-8980-945X]{Gang Zhao}
\altaffiliation{CAS Key Laboratory of Optical Astronomy, National Astronomical Observatories, Chinese Academy of Sciences,Beijing 100101, China}
\altaffiliation{School of Astronomy and Space Science, University of Chinese Academy of Sciences, Beijing 100049, China}
\author[0000-0002-8442-901X]{Yuqin Chen}
\altaffiliation{CAS Key Laboratory of Optical Astronomy, National Astronomical Observatories, Chinese Academy of Sciences,Beijing 100101, China}
\altaffiliation{School of Astronomy and Space Science, University of Chinese Academy of Sciences, Beijing 100049, China}

\begin{abstract}
We search for metal-rich Sausage-kinematic (MRSK) stars with $\feh > -0.8$ and
$-100<\Vphi<50\, \kmprs$ in LAMOST DR5 in order to investigate
the influence of the Gaia-Sausage-Enceladus (GSE) merger event on the Galactic
disk. For the first time, we find a group of low-$\alpha$
MRSK stars, and classify it as a metal-rich tail of
the GSE galaxy based on the chemical and kinematical properties.
This group has slightly larger $\Rapo$, $\Zmax$ and $\Etot$ distributions
than a previously-reported high-$\alpha$ group.
Its low-$\alpha$ ratio does not allow for an origin resulting
from the splash process of the GSE merger event, as is proposed to explain
the high-$\alpha$ group.
A hydrodynamical simulation by Amarante et al. provides a
promising solution, in which the GSE galaxy is a clumpy Milky-Way analogue
that develops a bimodal disk chemistry. This scenario explains the
existence of MRSK stars with both high-$\alpha$ and low-$\alpha$ ratios
found in this work. It is further supported by
another new feature that a clump of MRSK stars
is located at $\Zmax=3-5$ kpc, which corresponds to
the widely adopted disk-halo transition at $|Z|\sim4$ kpc.
We suggest that a pile-up of MRSK stars at $\Zmax$
contributes significantly to this disk-halo transition, an interesting imprint
left by the GSE merger event. These results also provide an important implication
on the connection between the GSE and the Virgo Radial Merger.
\end{abstract}

\keywords{Galactic halo; Spiral arms and galactic disk; Chemical composition and chemical evolution; Kinematics, dynamics, and rotation}



\section{INTRODUCTION}
It has building evidence that the Galaxy experienced several major
merger events and left imprints on the stellar halo, in the forms
of streams and overdensities in spatial position
and/or clumps and groups in integrals of motion (e.g., energy, angular
momenta, actions). Two famous mergers are the recent
Sagittarius (Sgr) and the ancient Gaia-Sausage-Enceladus (GSE) dwarf
galaxies. The former is witnessed by the existence of leading and
trailing arms in a large sky coverage \citep{Ibata94}, while
the latter was discovered as the so-called ``Gaia-Sausage" structure
in the radial versus azimuthal velocity space \citep{Belokurov18,Myeong18}
without any dense stream detected so far.
It is proposed that the GSE galaxy collided nearly head-on with
the Galaxy and was disrupted about $6-11$ Gyr ago, leaving a giant cloud of
intermediate-metallicity ($\feh \sim -1.3$) stars on highly radial orbits
\citep{Helmi18,Haywood18,Fattahi20}.
Tracing these events and searching for their imprints are
our long-term goals in order to understand the building blocks of
the Galactic halo, and perhaps also the Galactic disk,
within the context of $\Lambda$CDM cosmology.

As a recent merger with the Galaxy at a relatively low inclination, the debris of Sgr
are found all over the sky
with Galactocentric distances ranging from  15 kpc to
130 kpc \citep{Majewski03,Belokurov14,Sesar17}.
Being on a high eccentricity orbit, the ancient GSE merger left its debris
in a limited region on the sky within 20-30 kpc \citep{Iorio19},
and delivered large amounts of stars
into the solar neighborhood \citep{Belokurov18,Myeong19}.
Interestingly, \cite{Gallart19} suggested that low-$\alpha$ stars found
by \cite{Nissen10} within 0.3 kpc of the Sun
are also debris of the GSE merger. The
Sgr merger event provides a dominant component
in the outer halo ($R>30$ kpc), while the accreted component in the inner halo (and the
Solar neighborhood) mainly comes from the GSE merger
\citep{Naidu20}.

Meanwhile, \cite{Donlon19} suggested that the GSE merger may be
the same as the ``Virgo Radial Merger (VRM)," which includes the Virgo Overdensity,
the Virgo Stellar Stream, and many halo streams. They claimed that the
Virgo Overdensity, the Hercules-Aquila Cloud and
the Eridanus-Phoenix overdensity are probably
unmixed portions of the GSE tidal debris. A systematic search for
this connection has not been carried out and will become possible
with the launch of the future 2-m Chinese Space Survey Telescope (CSST) project \citep{Zhan11} taking into account
of its advantage
of both deep field and high spatial resolution observations.
Since stars belonging to the same structure share similar
abundance patterns \citep{Freeman02},
the chemical mapping of halo stars clustering in both spatial position and
integrals of motion by the CSST project will open the door to reconstruct
the Galactic past merging history.

Last but not least, it is proposed that the GSE merger event perturbed
the proto-disk and produced
a special metal-rich halo-like population in the biggest splash
\citep{Belokurov19}, which heats disk stars up to the Galactic halo.
In support for this scenario, a low angular momentum ($\Vphi<100 \kmprs$)
stellar population with $\feh>-1$ was
found by \cite{Bonaca17}, and was classified as
a ``heated thick disk", rather than an ``in-situ halo."
Later, more works \citep{Haywood18,DiMatteo19,Amarantea20} reported
the existence of Sausage-kinematic stars with thick-disk chemistry.
They were named ``splash" stars and become important
imprints of the GSE merger event on the Galactic disk.

In this paper, we aim to search for additional imprints of the GSE merger event
using the LAMOST low-resolution survey, in combination with Gaia DR2,
based on the chemical and kinematical properties of disk-metallicity stars.
Since the LAMOST survey provides the largest dataset of stars in
the Galactic disk,
it is favorable to investigate the influence of the GSE accretion
event on the evolution of the Galactic disk. It is expected that more
``splash" stars with disk chemistry can be found so that new
merging imprints may be detected from a good statistic analysis on big data.
Meanwhile, it is interesting to probe if there is
a metal-rich tail of the GSE galaxy extending from the main component
at $\feh \sim -1.3$, which will provide an important implication for its
connection with the VRM because the Virgo Overdensity
is metal rich and extends right into the Galactic disk.

\section{Selection of star samples}
The first stage of the LAMOST spectroscopic survey \citep{Zhao06,Zhao12,Cui12,Deng12}
was carried out in a low-resolution (R$\sim$2000) mode with a wide
spectral wavelength coverage of $3800-9000\,\AA$.
Based on abundances from APOGEE DR14 by \cite{Ting19},
\cite{Xiang19} developed a data-driven neural network and
derived stellar parameters and $\afe$ abundances for 8,162,566 stars
from LAMOST DR5. This is an internally consistent
dataset (hereafter DDPayne-LAMOST DR5) and thus becomes a good
sample for a statistic study.

From this catalog, we select stars with $\feh >-1.5$ and require
the errors in both $\feh$ and $\afe$ to be less than 0.1 dex.
We divide the dataset into three samples according to the gravity range,
$2.6<logg<3.3$ (LGB, low red-giant branch), $2.2<logg<2.6$ (RC, red clump)
and $0.0<logg<2.2$ (UGB, upper red-giant branch). With this division,
the analysis is based on more internally-consistent abundances within each
sample than in the whole dataset by reducing the
dependence of abundance on stellar parameters. Meanwhile, results from different samples can be compared and checked independently to draw a reliable conclusion. We do not include dwarf stars
from DDPayne-LAMOST DR5
because their abundances are less reliable
due to the uncertainty in the trained dataset of APOGEE DR14
\citep{Zasowski19}.

APOGEE (Apache Point Observatory Galactic Evolution Experiment)
\citep{Majewski17} is a medium-high resolution (R$\sim$22,500)
spectroscopic survey in the near-infrared spectral range
(H band, 15700-17500\AA). The most recent version is
APOGEE DR16, which provides more reliable parameters and
elemental abundances than APOGEE DR14 thanks to the updated pipeline {\rm ASPCAP} \citep{Garca16} using a grid of only MARCS stellar
atmospheres \citep{Gustafsson08} and a new H-band line list \citep{Smith20}.

We cross-match DDPayne-LAMOST DR5 with APOGEE DR16
and carry out calibrations of $\feh$ and $\afe$ for the selected three
samples. We limit stars with signal-to-noise larger than 80
and a good fitting quality of $\chi_{ASPCAP} < 25$ from APOGEE DR16 to get reliable abundances
for calibration. Fig.~1 shows the comparison of $\feh$, $\afe$
and calibrations by linear fitting between the two datasets for the three samples.
Corrections for $\feh$ are applied to
the three samples of DDPayne-LAMOST DR5 based on the individual
calibration of each sample so that the results from
DDPayne-LAMOST DR5 and APOGEE DR16 can be compared when necessary.
We do not correct $\afe$ ratios because there are
significant scatters in the calibrations, and we prefer to have internally
consistent $\afe$ within each sample from DDPayne-LAMOST DR5 for separating the thin and thick
disks in the following analysis.

We make use of distances from the StarHorse catalog for LAMOST
DR5 by \cite{Queiroz20} with the relative error in distance
less than 10\%. Radial velocities with errors less than $10 \kmprs$
are taken from LAMOST DR5 and proper motions with errors less than $0.25\,mas/yr$ are from {\it Gaia} DR2 \citep{Babusiaux18}.
We check the quality of the {\it Gaia} astrometry
with the renormalized unit weight error of $RUWE<1.44$ \citep{Lindegren18}.
Spatial velocities and orbital parameters are calculated based
on the publicly available
code {\it Galpot}. We employ the default potential of {\it MilkyWayPotential}
provided by \cite{McMillan17}, the solar position of
$R=8.2$ kpc, and the circular
speed $V_{c} = 233.1 \kmprs$ \citep{McMillan11}.
The peculiar velocity of the Sun is ($U_{\odot}, V_{\odot}, W_{\odot}) = $ (11.1, 12.24, 7.25) km/s \citep{Schonrich10}.

With abundances and velocities available, we select stars
with $-100<\Vphi<50 \kmprs$ and $\feh>-0.8$ as our targets
based on the characteristics of the Gaia-Sausage structure,
and we name them ``metal-rich Sausage-kinematic (MRSK)" stars in this work.
The choice of the upper limit of $50 \kmprs$ aims to
avoid stars at $\Vphi \sim 100 \kmprs$, which is
the transition between the
lower boundary of the thick disc and the upper
boundary of the splash box \citep{Belokurov19}.
We adopt a lower limit of $-100 \kmprs$ in order to include
more halo-kinematic disk-metallicity stars. This selection will not
include stars from the Sequoia dwarf galaxy, which has
retrograde orbits with $\Vphi<-100 \kmprs$ and is relatively metal poor
with $\feh < -1.5$ as shown in \cite{Belokurov19}.

Fig.~2 shows the $\VR$ versus $\Vphi$ diagrams and the contour maps
of $\feh$ versus $\afe$ for all stars
with $\feh>-0.8$ and $R>5$ kpc in the three samples. The latter limit is adopted
to avoid stars from the bar/bulge region ($R<5$ kpc). Within the two red
lines, the existence of MRSK stars is obvious.
For normal disk stars with $\feh > -0.8$ and $\Vphi>150 \kmprs$,
the separation between the
high-$\alpha$ thick disk and the low-$\alpha$ thin disks
is shown by the red line of $\afe=-0.08\feh+0.12$ in the $\feh$
versus $\afe$ diagram.
This line will be adopted to pick out
low-$\alpha$ MRSK stars from the high-$\alpha$ group in
the following analysis.
Here we try to adopt a strict criterion to select
the low-$\alpha$ group by lowering the red line close to the boundary
of the low-$\alpha$ contours for normal thin-disk stars.

\begin{figure*}
\centering
\includegraphics[scale=1.0]{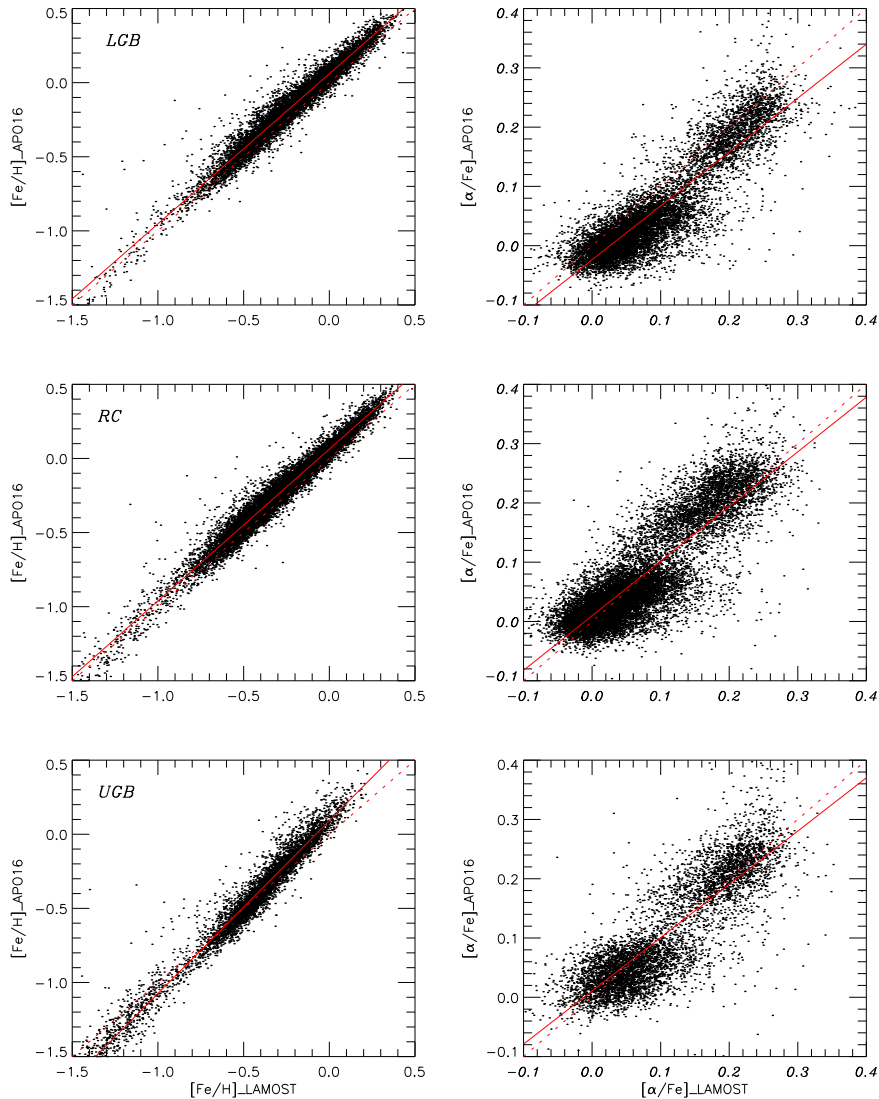}
\caption{Comparison of $\feh$ and $\afe$ between DDPayne-LAMOST DR5 and APOGEE DR16 for the three samples (LGB, RC and UGB). Red dash lines are the one-to-one relations and solid lines are calibrations derived from linear fits to the data.}
\label{f1}
\end{figure*}

\begin{figure*}
\centering
\includegraphics[scale=1.0]{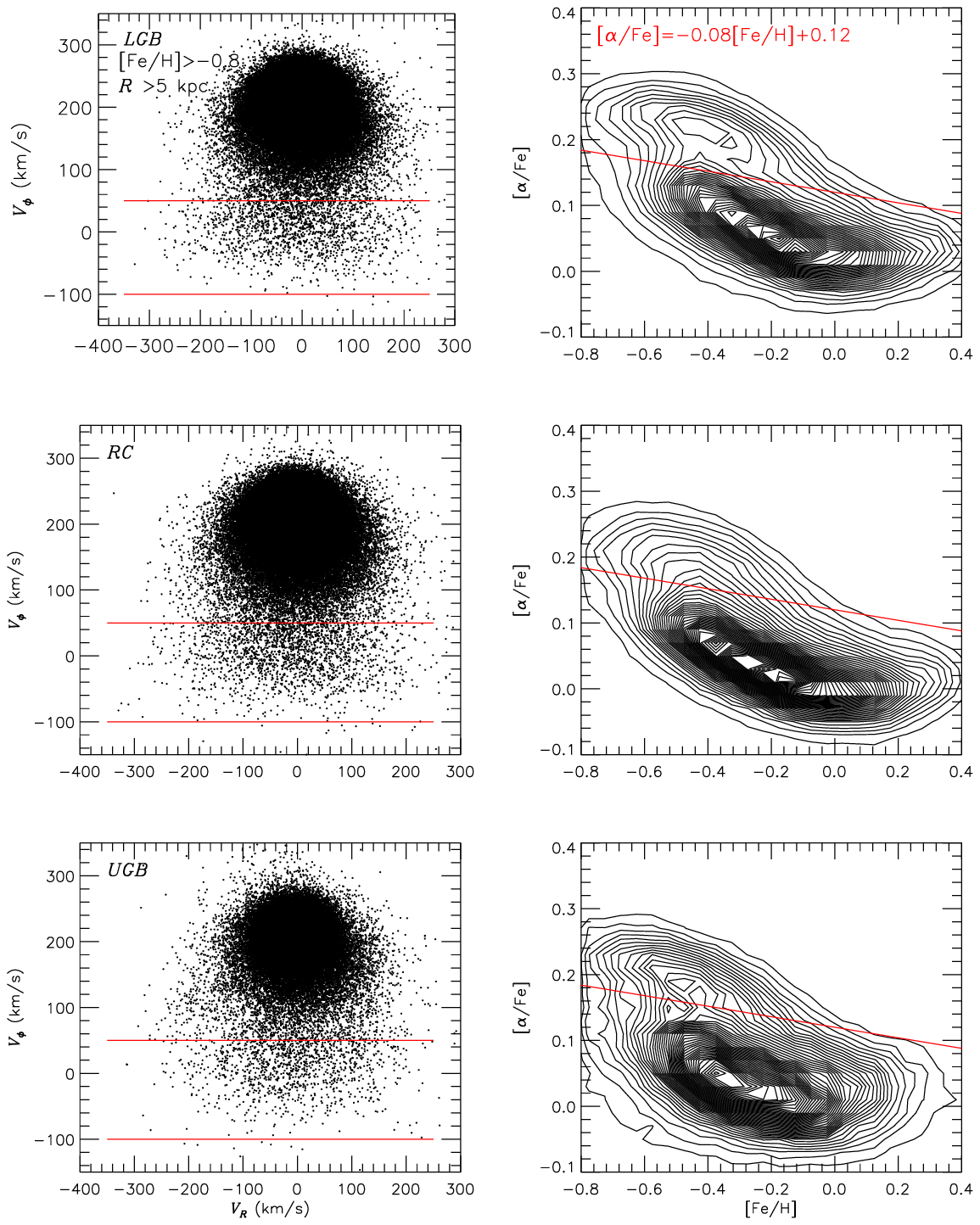}
\caption{Left: $\VR$ versus $\Vphi$ diagram for stars with $\feh>-0.8$ and $R>5$ kpc in the three samples. The two red lines show the definition of metal-rich Sausage-kinematic (MRSK) stars within $-100 <\Vphi< 50 \kmprs$. Right: Contour maps of $\feh$ versus $\afe$ for the three samples. Red lines are defined to separate between the thin and thick disks, which will be used to pick out the low-$\alpha$ group from the high-$\alpha$ group among MRSK stars.}
\label{f2}
\end{figure*}

\section{Two groups of MRSK stars with high-$\alpha$ and low-$\alpha$ ratios}
Fig.~3 shows the $\feh$ versus $\afe$ diagrams for MRSK
stars ($-100 <\Vphi< 50 \kmprs$ and $\feh > -0.8$)
in the three samples.
The high-$\alpha$ group above the red line is expected to be
the heated thick disk population, consisting of
so-called ``splash" stars, due to the massive merger event of the GSE galaxy
as suggested by \cite{Belokurov19}.
To our interest, a significant fraction of
MRSK stars have low-$\alpha$ ratios, and they exist in
three different samples. The low-$\alpha$ group extends to a
higher metallicity of $\feh\sim0.0$ (or above) than the
high-$\alpha$ group in each sample.
The existence of low-$\alpha$ MRSK stars is not previously
reported in the literature, and there are totally 1534 such stars
in the three samples. We check that their Sausage-kinematic
characteristics persists when we adopt the distance
by inverting the parallax of {\it Gaia} DR2.
It seems unlikely that errors in $\afe$ and/or kinematics
can produce similar distributions in the $\feh$ versus $\afe$ diagram
for three different samples. Therefore, we suggest that this population
of low-$\alpha$ MRSK stars is real.

As an independent check, we investigate if the low-$\alpha$ MRSK group
exists in the APOGEE survey by applying the same procedure
to APOGEE DR16 for stars in the gravity range of $0<logg<3.8$.
We adopt an extra criterion of $DEC > -10 \deg$, as similar as
the sky coverage of the LAMOST survey, and with the purpose to avoid stars
from Magellanic Clouds and Sagittarius Streams in the southern sky.
Due to the small number of this population, we do not divide
this dataset into three samples (LGB, RC and UGB) accordingly.
As seen in Fig.~4, MRSK stars in APOGEE DR16 show both
high-$\alpha$ and low-$\alpha$ ratios with an obvious gap,
according to the same separation line as that of DDPayne-LAMOST DR5.
The majority of MRSK stars belong to the high-$\alpha$ group, but
the low-$\alpha$ group (consisting of 410 stars) is significant.
Also, the low-$\alpha$ group in this sample
has an similar extension toward super solar metallicity as in DDPayne-LAMOST DR5.
If the systematic shift of 0.03 dex in $\afe$ to a lower value
in APOGEE DR16 (see Fig.~1) is taken into account, the star
number of the low-$\alpha$ group is slightly reduced (398 stars).
In addition, the separation between the high-$\alpha$ and
the low-$\alpha$ groups is more clear in APOGEE DR16
than in DDPayne-LAMOST DR5.
Thus, the existence of a low-$\alpha$ MRSK group is
significant and real in this independent dataset.

Finally, with close inspection on the data in \cite{Belokurov19}, we find that
the $\feh$ versus $\Vphi$ trend in their Figure 2 also shows a hint on
the existence of a metal-rich tail of the GSE galaxy. The distribution of stars
with $\Vphi \sim 0 \kmprs$ seems to cross the splash box and
extends to the metal-rich end until $\feh \sim 0$. Moreover, in
the $\VR$ versus $\Vphi$ diagram as coded with $\afe$,
Sausage-kinematic stars have both high-$\alpha$ and low-$\alpha$ ratios.
This supports the reality of the low-$\alpha$ MRSK group.

\begin{figure*}
\centering
\includegraphics[scale=1.0]{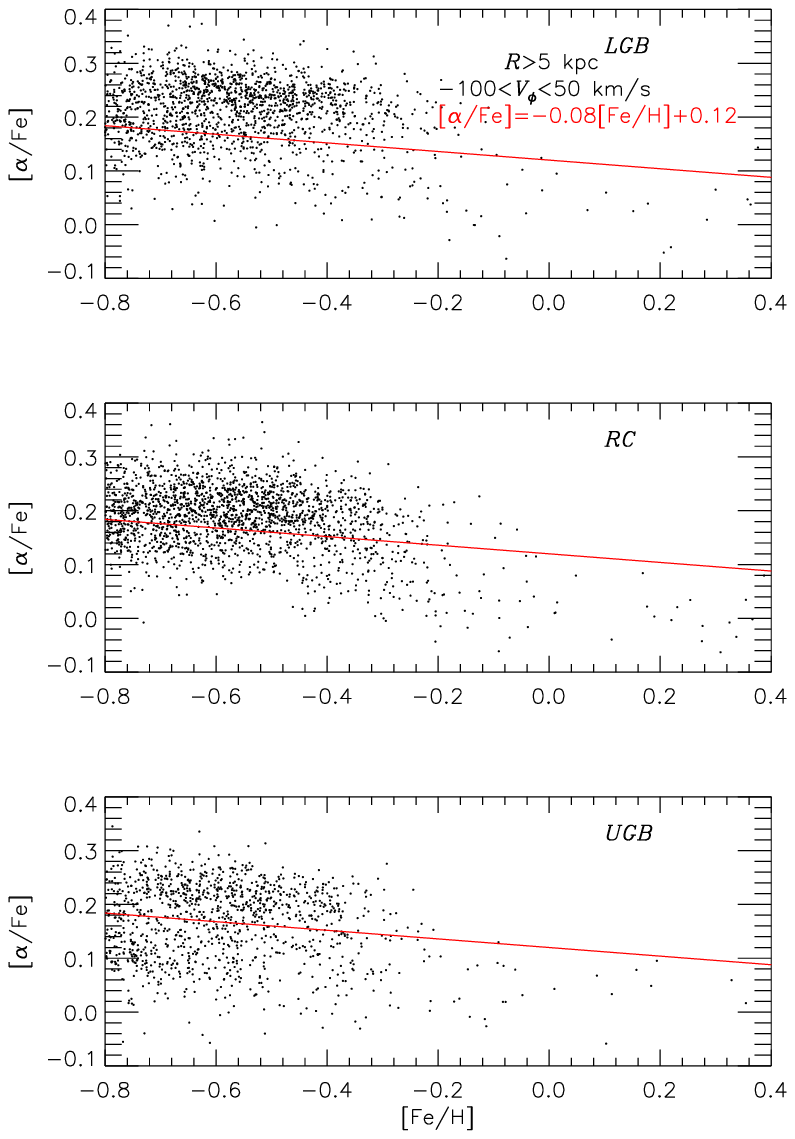}
\caption{$\feh$ versus $\afe$ diagrams for the three samples in DDPayne-LAMOST DR5 data. The red lines are the same as in Fig. 2.}
\label{f3}
\end{figure*}

\begin{figure*}
\centering
\includegraphics[scale=1.0]{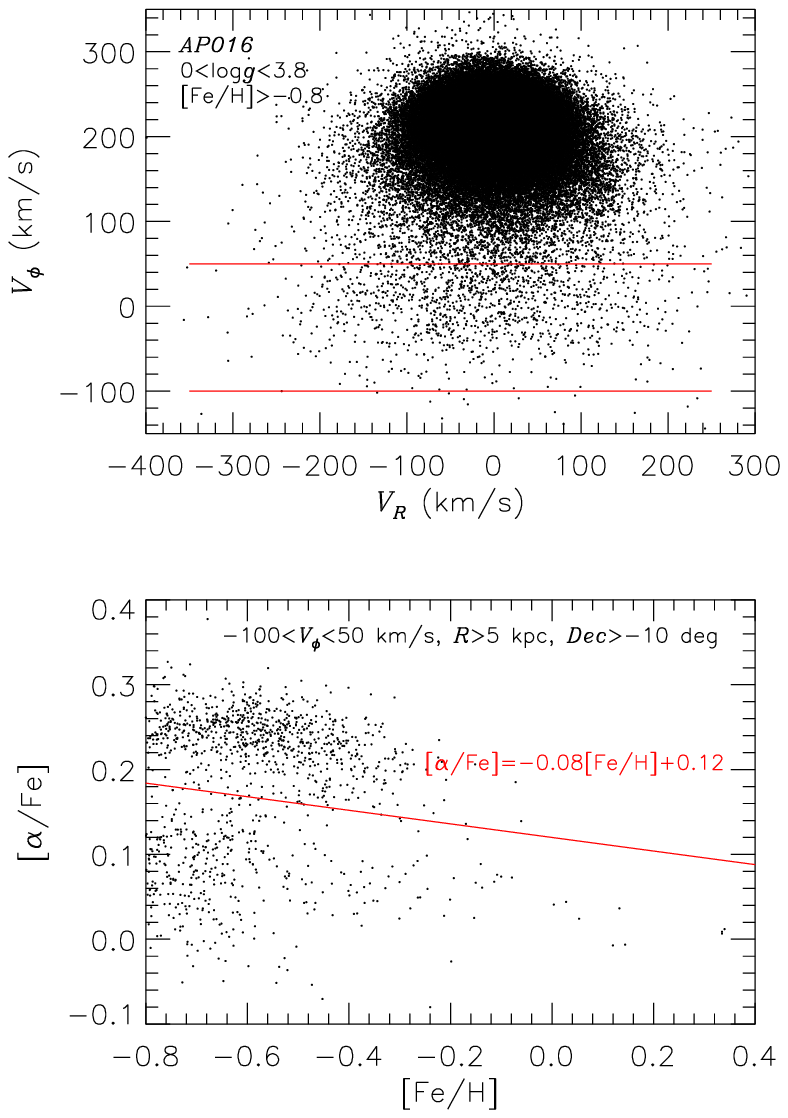}
\caption{Upper: $\VR$ versus $\Vphi$ (upper) diagram for stars with $0.0 < \logg < 3.8$, $\feh>-0.8$ and $Dec>-10$ in APOGEE DR16. Lower: Contour map of $\feh$ versus $\afe$ for MRSK stars (with $\feh>-0.8$ and $-100 < \Vphi < 50 \kmprs$) in APOGEE DR16. Lines are the same as in Fig.~2.}
\label{f4}
\end{figure*}

\section{Galactic locations and orbital properties between
high-$\alpha$ and low-$\alpha$ MRSK stars}
\subsection{The $R$, $|Z|$, $\Rapo$ and $\Zmax$ distributions}
The comparison of Galactic locations, in terms
of $R$ and $|Z|$, between the high-$\alpha$ and
low-$\alpha$ groups for the three samples is shown in Fig.~5.
The distributions of the high-$\alpha$
group are reduced by a factor of three in order to have similar peaks
as the low-$\alpha$ group for comparison. Generally,
there is no obvious difference in both $R$ and $|Z|$ distributions
between the high-$\alpha$ and low-$\alpha$ groups for the LGB and
the RC samples. As the luminosity increases (and includes more
distant stars) from the LGB/RC to UGB samples,
we would expect to observe more
old high-$\alpha$ stars than young low-$\alpha$
stars in the UGB sample.
However, this is not the case
and we observe an opposite trend for the UGB sample.
If the low-$\alpha$ group belongs to the locally-born thin-disk population,
it would be located below the splashed high-$\alpha$ thick disk.
Moreover, the GSE merger event happened about 9 Gyr ago,
before the formation of the thin disk. Thus, it is unlikely that these
low-$\alpha$ MRSK stars belong to the young thin-disk population,
and were splashed up to the halo during the GSE merger event. We suggest that
its low-$\alpha$ ratio indicates an origin from the
accreted halo (i.e. the GSE galaxy itself),
rather than from the thin disk.

It is interesting to investigate how far the accreted low-$\alpha$
group would go in the Galaxy and how different in the maximum distance
as compared with the splashed high-$\alpha$ group. In Fig.~6, the histograms of $\Rapo$ and $\Zmax$
are shown for both groups in the three samples. Obviously, the low-$\alpha$
group shows a systematic shift toward a
larger $\Rapo$ by $1-2$ kpc. Specifically, the LGB sample shows
a peak at 8.5 kpc for the high-$\alpha$ group and at 10.5 kpc
for the low-$\alpha$ group despite of their similar $R$ and $|Z|$
distributions. The $\Rapo$ distributions of the low-$\alpha$
group have wide peaks around 10 kpc for RC and UGB samples, extending
up to 15 kpc. The $\Zmax$ distributions generally have peaks around
4 kpc for both groups but the low-$\alpha$ group has extensions toward
higher values in the three samples. These distributions show that
the low-$\alpha$ group is not from the thin disk, which can not
reach a vertical distance as high as the typical thick disk and
would lie far below the splashed thick disk.
Finally, in view of the head-on collision
of the GSE galaxy with the Galaxy, similar peaks at 4 kpc in the
$\Zmax$ distributions
between the low-$\alpha$ and high-$\alpha$ groups may suggest
that they are related, and probably are connected by the same event.
We will discuss this possibility later.

\begin{figure*}
\centering
\includegraphics[scale=1.0]{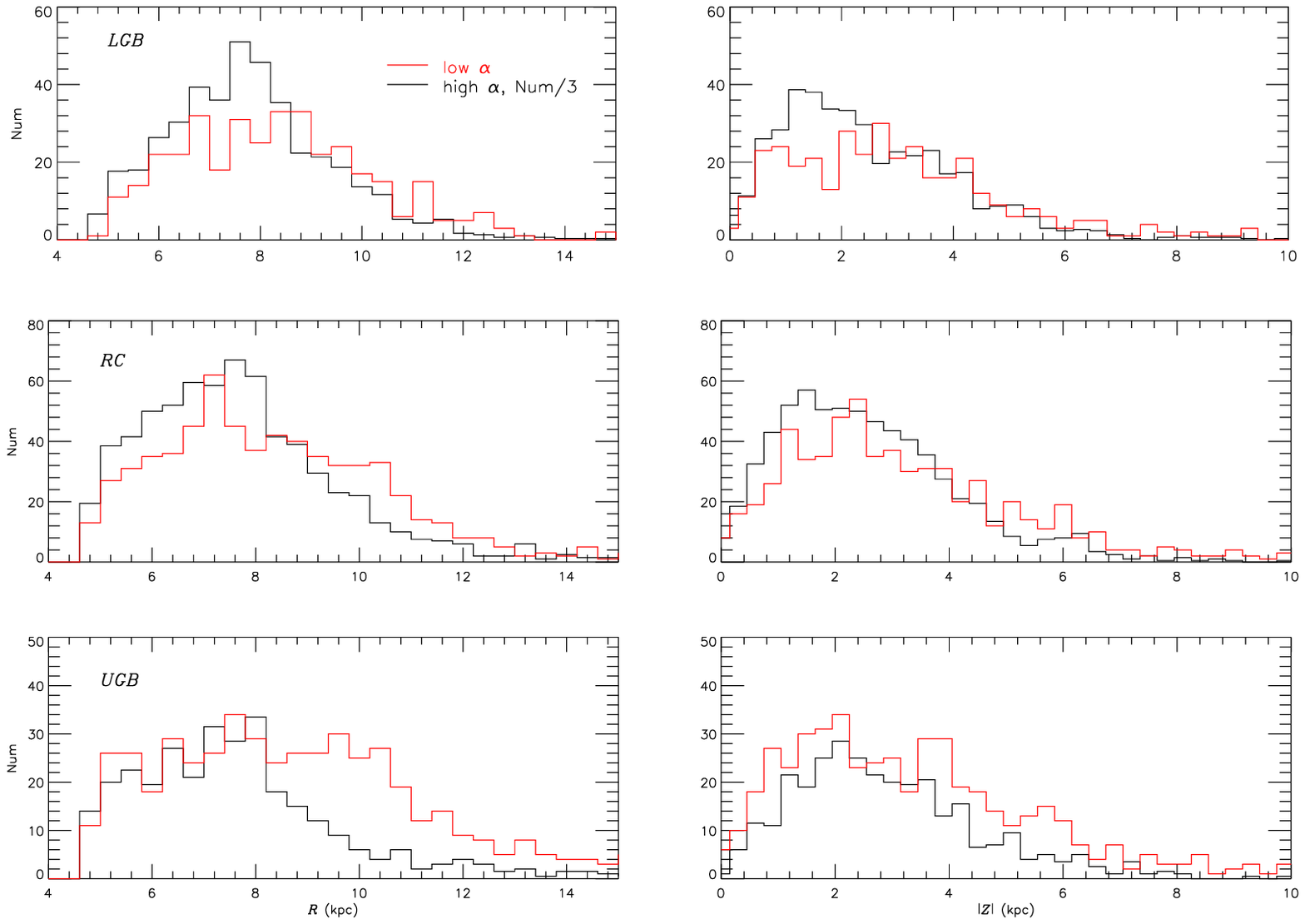}
\caption{Histograms of $R$ (left) and $|Z|$ (right) for MRSK stars with high-$\alpha$ (black) and low-$\alpha$ (red) ratios in the three samples of DDPayne-LAMOST DR5.}
\end{figure*}

\begin{figure*}
\centering
\includegraphics[scale=1.0]{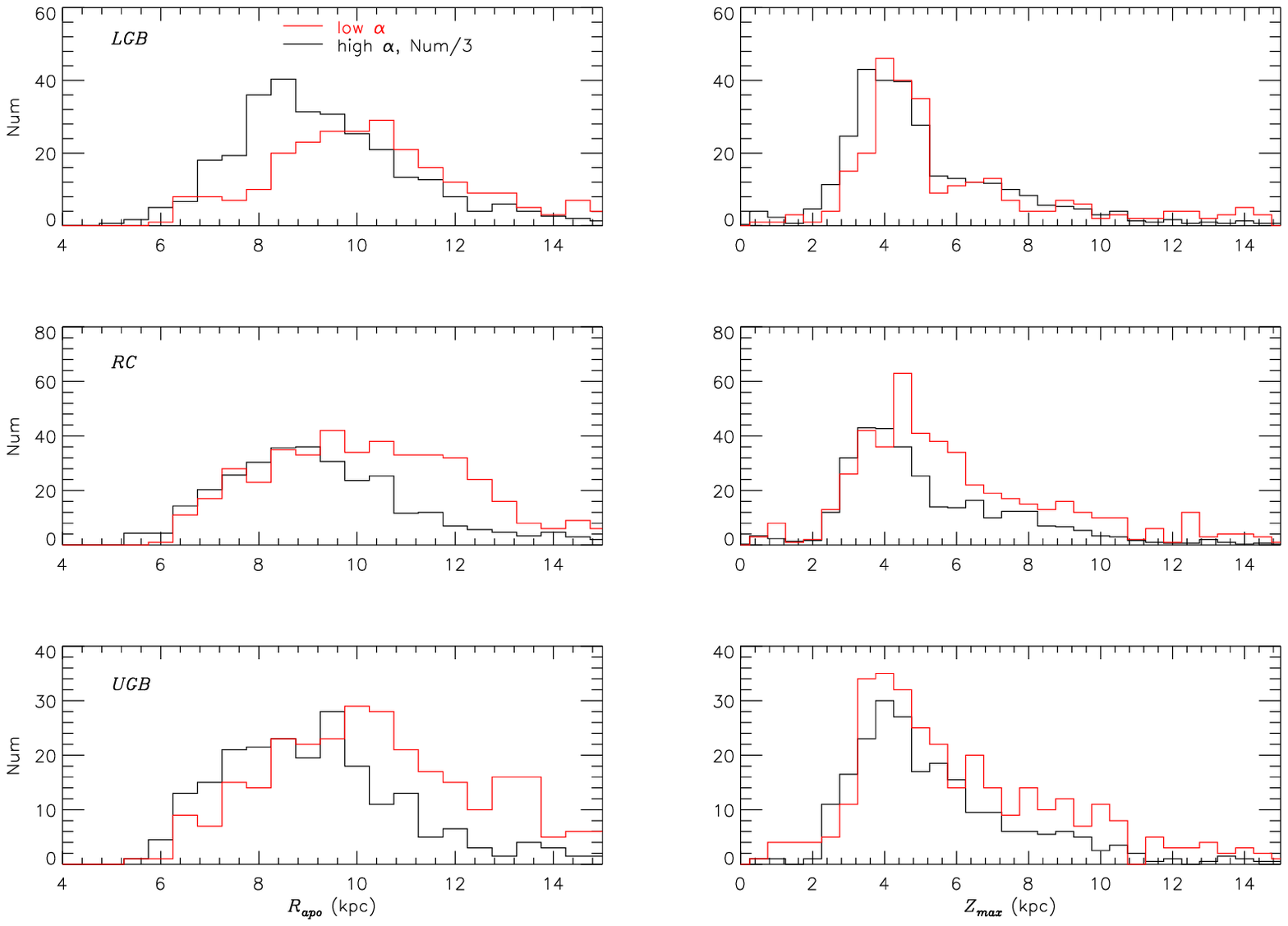}
\caption{Histograms of $\Rapo$ (left) and $\Zmax$ (right) for MRSK stars with high-$\alpha$ (black) and low-$\alpha$ (red) ratios in the three samples of DDPayne-LAMOST DR5.}
\label{f6}
\end{figure*}

\subsection{The $\Rm-\Zmax$ and $e-\Zmax$ planes}
As compared with its present location $R$, the mean Galactic distance
$\Rm$, as defined to be $(\Rapo+\Rperi)/2.0$, provides
more information on a star's origin.
Fig.~7 shows the $\Rm$ versus $\Zmax$ and the $e$ versus $\Zmax$ planes
for MRSK stars.  Interestingly, there are three sections
for both the high-$\alpha$ (small dots) and low-$\alpha$
groups. In particular, there is a clear gap along the black solid line
of $\Zmax=0.26\Rm+0.4$.
We note that, normal disk stars with
$\Vphi > 150 \kmprs$ are all located in the lowest panel of $\Zmax<2$ kpc.
That is, stars from the thin-disk population has
$\Zmax<2$ kpc and never show up at $|Z| > 2$ kpc.
This is a further support for the suggestion that the low-$\alpha$ MRSK group
does not belong to the thin disk population. Most MRSK stars from
the low-$\alpha$ group occupy the upper two sections with only a few exceptions.
The upper two sections are further separated by the dash line
of $\Zmax=0.9\Rm+1.4$, with blue open circles for
the middle sections and red circles for the upper sections.
The division lines are arbitrarily drawn by eye to separate the
three sections, and we compare the results within each section.
These features share similar characteristics to those in the $\Rapo-\Zmax$ diagram
of \cite{Amaranteb20}, which are caused by resonant trapping effects of the Galactic bar \citep{Moreno15}.
Interestingly, \cite{Schuster20} had shown that 3D star orbits of the high-$\alpha$ star G18-39 and
the low-$\alpha$ star G21-22 in \cite{Nissen10} are trapped by 2D resonant families V and IX of \cite{Moreno15}. Different orbital families in \cite{Moreno15} would reach different $\Zmax$, which explains the appearance of the three sections in Fig.~7.

The distribution of MRSK stars in three sections was previously reported by
\cite{Haywood18} in the $RD_{max}$ versus $\Zmax$ diagram (their Fig.~8)
for the high-$\alpha$ group. In this work, we found similar three-section
distributions for the low-$\alpha$ group in the $\Rm$ versus $\Zmax$ diagram.
This feature is unique to the GSE merger as a result from its high
eccentricity of $e>0.7$. As we limit stars
with $e>0.9$, five sections appear, while stars with $e<0.2$ show only one
section at the lowest part of $\Zmax<2$ kpc.
In view of this dependence, we show the $e$ versus $\Zmax$ diagrams for
stars in the three sections on the right panels of Fig.~7.
Interestingly, stars in the middle sections show a
clump at $\Zmax=3-5$ kpc along the blue line of $\Zmax=8.5e-3.5$, while the
upper sections show sparse distributions without any clump.
There are 82\%, 64\% and 66\% stars in the middle sections located
within 1 kpc (two blue dash lines) along the solid blue lines
for the three samples.

We investigate the distribution of the vertical velocity $V_z$
for stars within two blue dash lines. There is almost zero $V_z$
with a small dispersion of $45 \kmprs$, significantly lower than that
of $\sim 90 \kmprs$ for stars above the upper blue dashed line and
of $\sim 110 \kmprs$ for the Galactic halo.
In particular, the $V_z$ dispersion of $45 \kmprs$ is typical
for the thick disk. Therefore, these clump stars from
the accreted halo could be regarded as the thick disk in kinematics.
We thus propose a scenario that a pile-up of stars at $\Zmax
=3-5$ kpc by the GSE merger event contributes significantly
to (or may be responsible for) the division
between the thick disk and the halo at $|Z|\sim4$ kpc, widely adopted in
the literature \citep{Kinman11,Fernandez19}.
In this scenario, due to the negligible velocity in both $\Vphi$ and $V_z$,
these clump stars, with either low-$\alpha$ or high-$\alpha$ ratio, will stay
at the maximum vertical distance ($\Zmax$) for a longer time than
in other positions, which leads to a pile-up of stars at this distance.
Specifically, the pile-up at $\Zmax=3-5$ kpc (due to the GSE merger event)
produces an observational feature that there is a high density of
stars at $|Z|\sim4$ kpc, beyond which the star number decreases. Thus
this clump is regarded as the division between the thick disk and the halo.
This suggestion is inspired by \cite{Deason13} who proposed that
a rapid transition in structural properties of the Galactic stellar halo
at the break radius of 20-30 kpc \citep{Watkins09,Deason11}
is due to the pile-up in $\Rapo$ of the tidal debris from a small number
of significant mergers, e.g. the Sgr galaxy. Note that the intermediate-eccentricity orbit
of the Sgr merger event will lead to the pile-up of stars in $\Rapo$.
With high-eccentricity
stars, the GSE head-on merger event tends to cause a pile-up in $\Zmax$, rather than $\Rapo$.
It is important to point out
that both the low-$\alpha$ and
high-$\alpha$ groups contribute to the formation of
the high density of stars at $|Z|\sim4$ kpc, and
thus they could be related to the same merger event.
This suggestion is also consistent with the result by \cite{Helmi18} that the GSE
merger led to the formation of the Galactic inner halo and the thick disk.

\begin{figure*}
\centering
\includegraphics[scale=1.0]{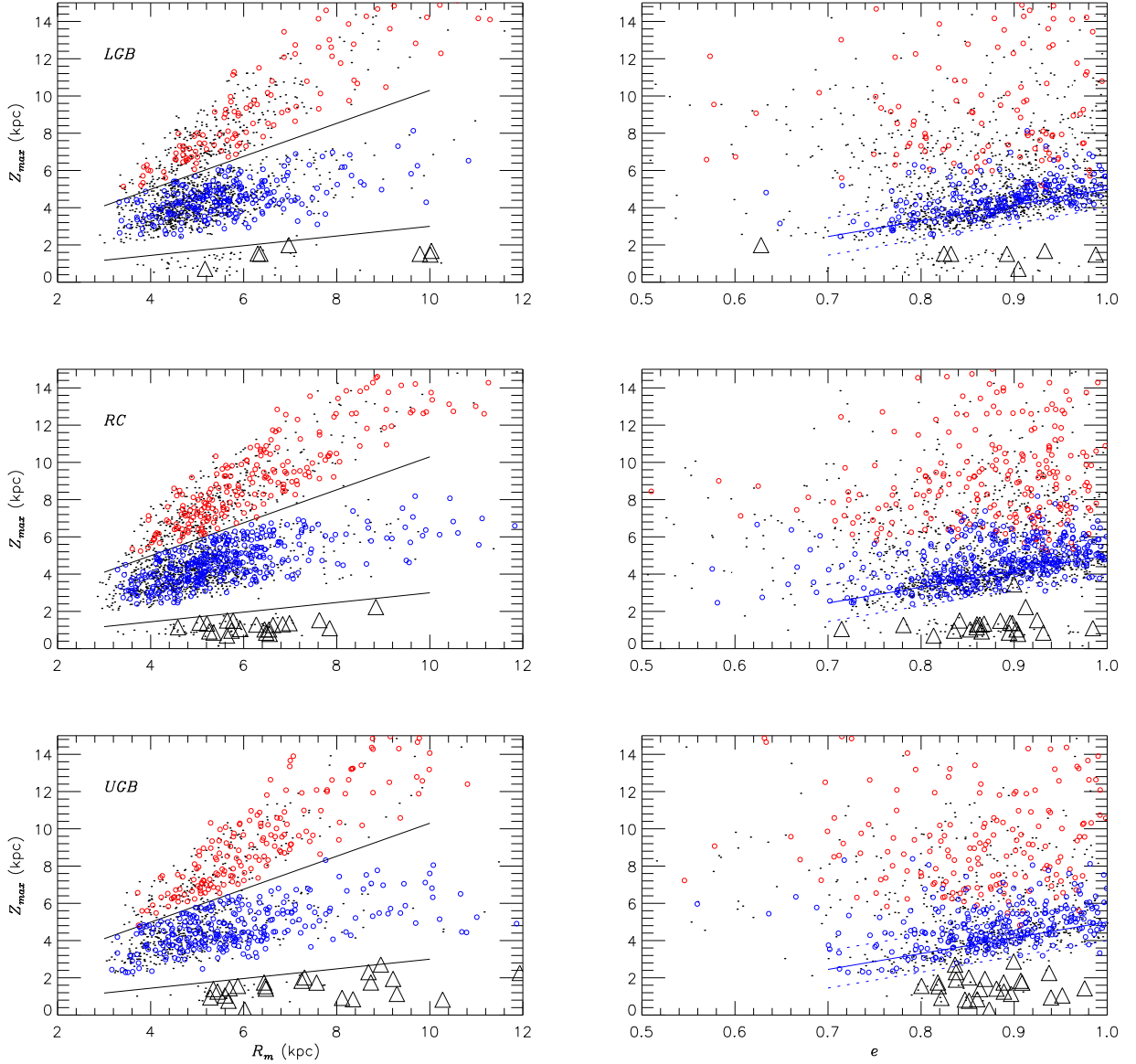}
\caption{$\Rm$ versus $\Zmax$ (left) and $e$ versus $\Zmax$ (right) diagrams for MRSK stars with high-$\alpha$ (black dots) and low-$\alpha$ (others) ratios in the three samples of DDPayne-LAMOST DR5. Low-$\alpha$ MRSK stars in the three sections are shown by black triangles (lowest), blue circles (middle) and red circles (upper).}
\label{f7}
\end{figure*}

\subsection{The $\Lz$ versus $\Etot$ diagram}
Fig.~8 shows the $\Lz$ versus $\Etot$ diagrams for
the low-$\alpha$ group in the three sections, as well as the
the high-$\alpha$ group (small dots). With significant overlapping,
the lowest section lies both sides, and the upper section
distributes more close to the central part around $\Lz\sim 0$ and
has a higher energy than the middle section. Specifically,
the lowest parts of Fig.~8 are occupied by stars in the middle
sections (blue circles) and the highest parts are mainly stars
from the upper sections (red circles). However, the overlapping
of both groups is dominated at $\Etotn \sim -1.7\, km^2s^{-2}$.
Therefore, we favor for the suggestion that no obvious
difference between the low-$\alpha$ and high-$\alpha$ groups is found
in this diagram.

The energy of MRSK stars is above the low boundary of the GSE
galaxy at $\Etotn \sim -1.9 \, km^2s^{-2}$, although its metal poor
component lies at $\Etotn \sim -1.5 \, km^2s^{-2}$ according to
\cite{Horta20}. In addition, the energy of the GSE galaxy is significantly
lower than the Sgr galaxy at $\Etotn \sim -1.0 \, km^2s^{-2}$
according to \cite{Naidu20} (see their Fig.~23).
Due to its higher energy, the Sgr merger event
has significant influence on the Galactic halo, since the majority of
member stars can reach up to the outer Galaxy (beyond $20-30$ kpc).
The lower energy of the GSE merger event makes it contribute
significantly to the formation of the Galactic disk, for example,
producing a clump of stars at $\Zmax=3-5$ kpc.

\begin{figure*}
\centering
\includegraphics[scale=1.0]{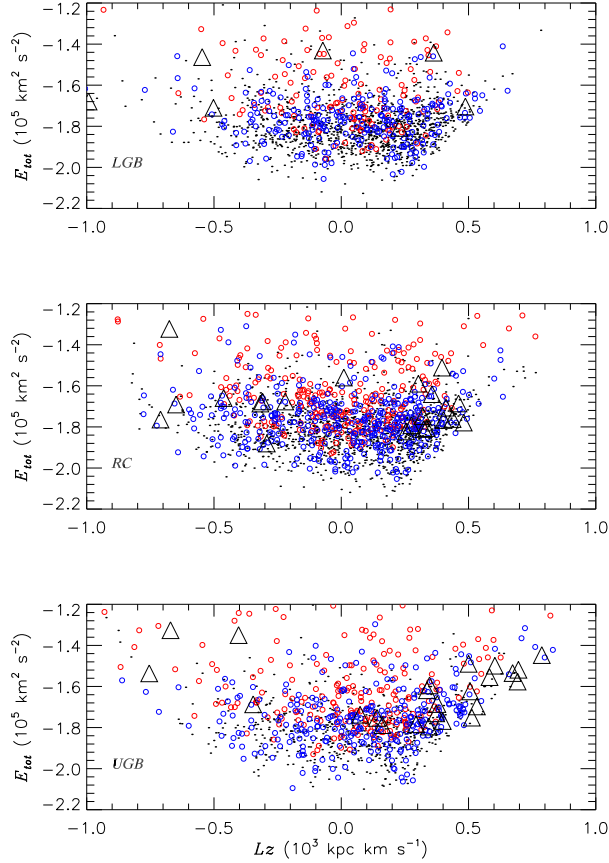}
\caption{$\Lz$ versus $\Etot$ diagrams for MRSK stars with high-$\alpha$ (black) and low-$\alpha$ (red) ratios in the three samples of DDPayne-LAMOST DR5. Symbols are the same as in Fig.~7.}
\label{f8}
\end{figure*}

\section{The origin of the low-$\alpha$ MRSK group}
\subsection{Connecting the low-$\alpha$ group to the GSE galaxy}

It is well known that Magellanic Clouds (MCs) and Sgr streams have metal rich
components. In APOGEE DR16, it is possible to compare
the low-$\alpha$ MRSK stars with member stars from MCs and Sgr
in the $\feh$ versus $\afe$ diagram. For this purpose,
stars with ``programnames" of ``magclouds" and ``Sgr" in
APOGEE DR16 are picked out, and we limit stars
within the range of $-100<\Vphi<50 \kmprs$, i.e. Sausage-kinematics.
In Fig.~9, these low-$\alpha$ MRSK stars (red circles)
are compared with Sausage kinematic stars from MCs (green crosses) and Sgr
(blue squares), as well as the metal poor GSE members (black dots).
It shows that the low-$\alpha$ MRSK group is
distinct from MCs at $\feh < -0.8$, but has
some overlapping at $-0.8<\feh < -0.4$. Meanwhile, GSE has higher $\afe$
than Sgr for the whole metallicity range of $-1.4<\feh < -0.2$, and
thus the low-$\alpha$ MRSK group does not belong to the Sgr galaxy.
In short, the low-$\alpha$ MRSK stars in APOGEE DR16 are distinct from the member stars of MCs and Sgr. They belong to the GSE galaxy as addressed in Sect. 4.1.

In Fig.~10, stars with $\feh > -1.5$ and $-100<\Vphi<50 \kmprs$
in DDPayne-LAMOST DR5 data are shown in the $\feh$ versus $\afe$
diagrams with the same symbols and colors for the three sections as in Fig.~7.
Three features in Fig.~10 are interesting and useful for understanding the origin of
the low-$\alpha$ group.
Firstly, there is no difference in $\afe$ among the three sections for
the low-$\alpha$ group. All of them belong to the same
population. Secondly, the contour maps show the main component of
the GSE galaxy at $\feh \sim -1.3$ and the second component
at $\feh \sim -0.5$ with a low-density gap at $\feh \sim -0.8$.
Thirdly, the second component becomes weaker from the LGB, through
the RC, to the UGB samples, as
the weak low-$\alpha$ tail becomes more
significant.
In the UGB sample, there is a clear trend of $\afe=-0.20\feh-0.05$
that connects the low-$\alpha$ MRSK group with the main GSE component
at $\feh \sim -1.3$. This connection provides a strong support for
the suggestion that the low-$\alpha$ MRSK group in DDPayne-LAMOST DR5 is
the metal-rich tail of the GSE galaxy.
By shifting the red line upward by 0.05 dex, the trend of
$\afe=-0.20\feh$ indicates the existence of
the metal-rich tail of the GSE galaxy in the LGB and RC samples.

\begin{figure*}
\centering
\includegraphics[scale=1.0]{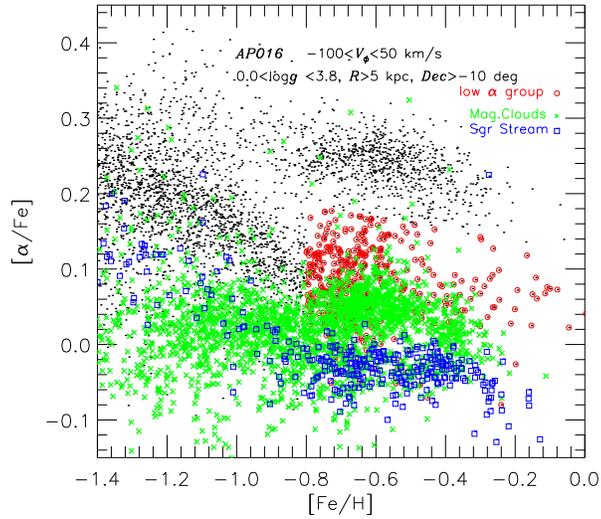}
\caption{$\feh$ versus $\afe$ diagrams for Sausage-kinematic stars with $\feh > -1.5$, $R>5$ kpc and $Dec > -10$ in APOGEE DR16. Sausage-kinematic stars from Magellanic Clouds (green crosses) and Sagittarius streams (blue squares) are overplotted.
}
\label{f9}
\end{figure*}

\begin{figure*}
\centering
\includegraphics[scale=1.0]{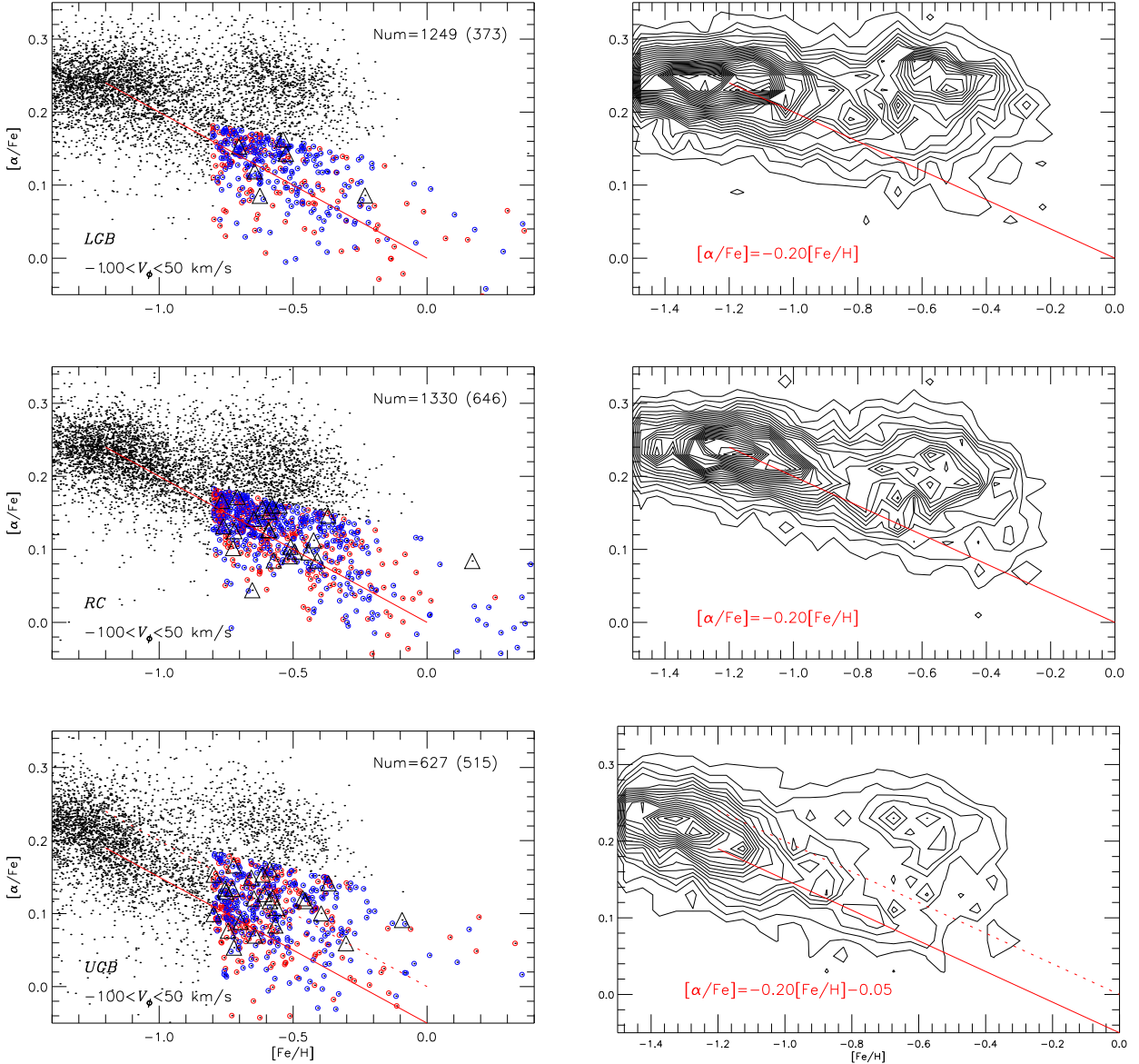}
\caption{$\feh$ versus $\afe$ diagrams and their contour maps for Sausage-kinematic stars with $\feh > -1.5$ and $R>5$ kpc in the three samples of DDPayne-LAMOST DR5. Symbols are the same as in Fig.~7.}
\label{f10}
\end{figure*}

\subsection{A scenario for the origin of the low-$\alpha$ MRSK group}
As described in Sect. 4, similar properties of the low-$\alpha$ and high-$\alpha$ MRSK groups in the $\Rm$-$\Zmax$ and $e-\Zmax$ diagrams indicate that
they probably result from the same process. However, the proposed
splash process by the GSE merger event for the origin of
the high-$\alpha$ group by \cite{Belokurov19} can not explain
the existence of the low-$\alpha$ group. Since $\alpha$ is an age-sensitive abundance \citep{Delgado19}, the
low-$\alpha$ ratio of this MRSK group indicates a young population, which is
not compatible with the splash process of the GSE merger event happened at an
ancient time.

Without a splash process, \cite{Amarantea20} developed a hydrodynamical
simulation, in which a clumpy Milky-Way-like analogue succeeds to produce
a bimodal disk chemistry with Sausage-kinematics, similar as our data.
In their Fig.~1, stars with $-100 <\Vphi< 50 \kmprs$ have metallicity
range from $\feh=-1.5$ to solar metallicity. Moreover,
both high-$\alpha$ and low-$\alpha$ stars exist in the disk metallicity range
of $\feh > -0.8$ (their Fig.~2). In particular, this scenario
does not need to exclude the GSE merger event. According to \cite{Mandelker14},
clumps in a galaxy can have an ex-situ origin associated to mergers.
It is possible that the GSE merger event had brought
the gas-rich cloud
into the Galaxy at an early time, and produced clumps that developed the
bimodal chemistry such as the high-$\alpha$ and low-$\alpha$ MRSK stars
in later evolution. This is a promising solution to the existence of
low-$\alpha$ MRSK stars, but it requires further study
to fit into the case of the GSE dwarf galaxy in more details.

\section{Summary and Conclusions}
A large sample of stars with disk metallicity in the LAMOST survey
provides us a good opportunity to find additional imprints of
the GSE merger event and to extend the search for its member stars
toward solar metallicity. With this purpose, metal-rich stars with Sausage-kinematics,
i.e. $\feh>-0.8$ and $-100 <\Vphi< 50 \kmprs$, in LAMOST DR5
are analyzed in spatial, chemical and kinematical planes.
We divide the stars into three samples (LGB, RC and UGB) so that our analysis is based on higher internally-consistent abundances within each sample of the low-resolution LAMOST survey.

A group of low-$\alpha$ MRSK stars is found, and it is thought to
be the metal-rich component of the GSE merger. In the UGB sample,
there is a continuous trend of $\afe=-0.20\feh-0.05$ that connects it
to the main body of the GSE galaxy at $\-1.3<\feh<-1.1$.
Moreover, it has larger $\Rapo$ (and $\Zmax$) distributions than
those of high-$\alpha$ MRSK stars, which excludes the possibility
to be the low-$\alpha$ thin disk population.
This is the first report on the existence of a metal-rich tail of
the GSE galaxy taking advantage of the large number of disk-metallicity stars in the LAMOST
survey. It provides important implication on the connection between the GSE galaxy and
the Virgo Radial Merger.

In the $\Rm$ versus $\Zmax$ plane, both the high$-\alpha$ and
low-$\alpha$ groups show three-section distributions
separated by $\Zmax = 0.26\Rm+0.4$ and $\Zmax = 0.90\Rm+1.0$ lines.
This feature is unique to the GSE galaxy due
to its high eccentricity of $e>0.7$. Most MRSK stars are located in the middle
and upper sections, while disk-metallicity stars with $\Vphi > 150 \kmprs$
limit within the lowest section of $\Zmax < 2$ kpc. According to \cite{Amaranteb20}, the three-section distributions in the $\Rm$ versus $\Zmax$ plane are caused by different orbital families presented in \cite{Moreno15}.

We find an interesting clump of MRSK stars at $\Zmax=3-5$ kpc,
along the line of $\Zmax=8.5e-3.5$ in the $e$ versus $\Zmax$ diagrams. Since this clump
has negligible velocity in $\Vphi$ and $V_z$, its stars would spend a longer time
at $\Zmax$ than in other positions, leading to a pile-up of MRSK stars
at $|Z|=3-5$ kpc. This clump corresponds to the widely-adopted
disk-halo transition at $|Z|\sim 4$ kpc, and thus is an interesting imprint left by
the GSE merger event. This is new evidence for the scenario by \cite{Helmi18} that
the GSE merger event led to the formation of the Galactic inner halo and the thick disk.

The splash process of the GSE merger event, as proposed to explain the high-$\alpha$ MRSK group, fails to account for
the low-$\alpha$ group because the ancient GSE merger event happened
before its formation.
Instead, a clumpy Milky-Way-like analogue in the hydrodynamical
simulation by \cite{Amarantea20} can produce a bimodal disk chemistry,
similar as our data, without the need for a splash process.
Based on this scenario, both high-$\alpha$ and low-$\alpha$ MRSK stars
in the LAMOST survey belong to the GSE galaxy.
The ancient GSE merger event had brought the gas-rich cloud
into the Galaxy at an early time, and produced clumps
that developed this bimodal chemistry for MRSK stars in
later evolution. This is a promising solution to the origin
of the low-$\alpha$ MRSK stars, but further study using simulations is
necessary to confirm this scenario.

\section*{Acknowledgements}
This study is supported by the National Natural Science Foundation of China (Grant Nos. 11988101, 11625313, 11890694), National Key R\&D Program of China (Grant No. 2019YFA0405502) and the 2-m Chinese Space Survey Telescope project. We would like to extend our sincere thanks to Dr. Sarah A. Bird for her kind help.

Guoshoujing Telescope (the Large Sky Area Multi-Object Fiber Spectroscopic Telescope, LAMOST) is a National Major Scientific Project has been provided by the National Development and Reform Commission. LAMOST is operated and managed by the National Astronomical Observatories, Chinese Academy of Sciences.\\
Funding for the Sloan Digital Sky Survey IV has been provided by the Alfred P. Sloan Foundation, the U.S. Department of Energy Oce of Science, and the Participating Institutions. SDSS-IV acknowledges support and resources from the Center for High-Performance Computing at the University of Utah. The SDSS web site is www.sdss.org.

SDSS-IV is managed by the Astrophysical Research Consortium for the Participating Institutions of the SDSS Collaboration including the Brazilian Participation Group, the Carnegie Institution for Science, Carnegie Mellon University, the Chilean Participation Group, the French Participation Group, Harvard-Smithsonian Center for Astrophysics, Instituto de Astrofsica de Canarias, The Johns Hopkins University, Kavli Institute for the Physics and Mathematics of the Universe (IPMU) / University of Tokyo, the Korean Participation Group, Lawrence Berkeley National Laboratory, Leibniz Institut f\"ur Astrophysik Potsdam (AIP), Max-Planck-Institut f\"ur Astronomie (MPIA Heidelberg), Max-Planck-Institut f\"ur Astrophysik (MPA Garching), Max-Planck-Institut f\"ur Extraterrestrische Physik (MPE), National Astronomical Observatories of China, New Mexico State University, New York University, University of Notre Dame, Observatario Nacional/ MCTI, The Ohio State University, Pennsylvania State University, Shanghai Astronomical Observatory, United Kingdom Participation Group,Universidad Nacional Autonoma de Mexico, University of Arizona, University of Colorado Boulder, University of Oxford, University of Portsmouth, University of Utah, University of Virginia, University of Washington, University of Wisconsin, Vanderbilt University, and Yale University.

\end{document}